# Mediated semi-quantum key distribution in randomization-based environment

Yen-Jie chen[1], Tzonelih Hwang[2], and Chia-Wei Tsai[3]

*[1,2]Department of Computer Science and Information Engineering, National Cheng Kung University, No. 1, University Rd., Tainan City, 70101, Taiwan, R.O.C.*

*[3]Department of Computer Science and Information Engineering, Southern Taiwan University of Science and Technology, No. 1, Nan-Tai Street, Yungkang Dist., Tainan City 710, Taiwan R.O.C.*

[1] mathmatic83@gmail.com

[2] hwangtl@csie.ncku.edu.tw (corresponding author)

[3] cwtsai676@stust.edu.tw

## Abstract

This paper proposes the first mediated semi-quantum key distribution in randomization-based environment with an untrusted third party (TP) who has the complete quantum capabilities to help two "classical" users to establish a secure key. The entanglement swapping between the collapsed qubits of Bell states and the Bell states is used to facilitate the design of the RSQKD protocol.

## 1. Introduction

In 1984, Bennett et al. [1] proposed the first quantum key distribution protocol. Afterward, numerous quantum key distribution protocols have been proposed [2-14]. However, the participants in these protocols are allowed to use all kinds of quantum devices (e.g., quantum memory, quantum generator, and so on.), which are still very expensive in realization so far.

To reduce the quantum burden on the participants, Boyer et al. [15,16] proposed the concept of semi-quantum key distribution (SQKD) protocols, in which the "quantum users" (or servers) usually have powerful quantum capabilities, Including generating quantum states (e.g., single photon and entangled quantum states), performing measurement (e.g., X-basis, Z-basis, or Bell measurement), storing qubits in quantum memory, and so on. However, the "classical users" can only have limited quantum capabilities including : (1) Z-basis measurement, (2) generating Z-basis qubits, (3) reflecting the qubits without any disturbance, (4) reordering the qubits by using different delay lines.

Based on the quantum operations the classical users own, Boyer et al. [15,16] proposed two kinds of semi-quantum key distribution (SQKD) protocols, namely the Measure-Resend SQKD protocol and the Randomization-Based SQKD protocol. In the Measure-Resend SQKD protocol, the classical users can only perform (1), (2) and (3) quantum operations to execute a protocol. On the other hand, the classical users can only perform (1), (3) and (4) operations in the Randomization-Based SQKD protocol. Afterward, various semi-quantum protocols and security analyse [15-22] have been proposed.

In 2015, Zou and Qiu et al. proposed another semi-quantum environment [22] named "Measurement-Free semi-quantum protocol" (MFSQP), in which the classical users are limited to perform (2), (3), (4) operations. In terms of performance, the MFSQP and Randomization-Based environment could have a better qubit efficiency than the Measure-Resend environment because they need less qubits.

To help two classical users to share a secret key between each other, a third-party (or a quantum user) is frequently introduced. Due to the use of quantum properties, the trustworthiness of the TP can be dishonest [23-28]. Based on this, Krawec et al. [29] and Liu et al. [30] proposed a Mediated Measure-Resend SQKD protocol and a Mediated Measurement-Free SQKD protocol in 2015 and 2018, respectively. However, there is still no Mediated Randomization-Based SQKD (MRSQKD) with a dishonest TP based on Randomization-Based environment.

In some situations, the users may not be able to do quantum generation, whereas it is easy for them to perform quantum measurement. In this case, the MR SQKD protocols may be useful for these users to establish session keys for secure communications. For example, the users in the BB84 [1] or BBM92 [47] can perform (1) Z-basis measurement, (2) X-basis measurement. If two users in BB84 or BBM92 want to negotiate a key between each other, then, the MRSQKD could be the best choice for them. Therefore, the purpose of this study is to propose a MRSQKD protocol with a dishonest TP, and then give a comparison to [29] and [30].

The rest of the paper is organized as follows. Section.2 introduces the entanglement swapping between the collapsed Bell state qubits and Bell states. Section.3 proposes the Mediated Randomization-Based SQKD protocol with a dishonest TP. Section 4 provides the security analyses and the comparison with the other protocols with same scenario. Finally, conclusions are given in Section 5.

## 2. The Entanglement correlations among collapsed Bell state qubits and reordered Bell states

In optical quantum mechanism, if one qubit in a Bell state is measured with Z-

basis, then the measured qubit will disappear and the other qubit in this Bell state will be collapsed into a Z-basis single qubit, named here as a **collapsed Bell state qubit**. This section will investigate Entanglement correlations among collapsed Bell state qubits and reordered Bell states after performing Bell Measurements.

Let $n$ Bell states in $\{|\phi^{\pm}\rangle, |\psi^{\pm}\rangle\}$, $n \in Z^+$, be prepared, and all the first and second qubits of Bell states form two ordered qubit sequences $S_1$ and $S_2$ respectively. Arbitrary $\frac{n}{2}$ positions in $S_1$ are measured with the Z-basis and then the remaining $\frac{n}{2}$ qubits are reordered to form $S_1'$; similarly, arbitrary $\frac{n}{2}$ positions in $S_2$ are measured with the Z-basis and then the remaining $\frac{n}{2}$ qubits are reordered to form $S_2'$. Bell Measurements are performed on both the *ith* elements of $S_1'$ and $S_2'$, $1 \leq i \leq \frac{n}{2}$, respectively, then various kinds of possible entanglement groups can be identified (see also Figure.1).

The first kind of groups is obtained while a Bell Measurement is performed on an original Bell state. The second kind of groups is the Entanglement Swapping while multiple Bell Measurements are performed on pairs of qubits from different Bell states. The third kind of groups is obtained while the Bell Measurement is performed on a pair of two collapsed Bell state qubits. The fourth kind of groups is obtained by performing somewhere within the entanglement groups two Bell Measurements on two special pairs of qubits, where each pair is formed by one collapsed Bell state qubit and one qubit from some other Bell state. Moreover, if the first collapsed Bell state qubit is from $S_1'$, then the other collapsed Bell state qubit is from $S_2'$, no matter how many Entanglement Swappings occur in-between them, and vice versa. Notice that the Group 4 in Figure 1 is the entanglement group after certain arrangement on the original entanglement group making the two collapsed Bell state qubits in the both ends of this group.

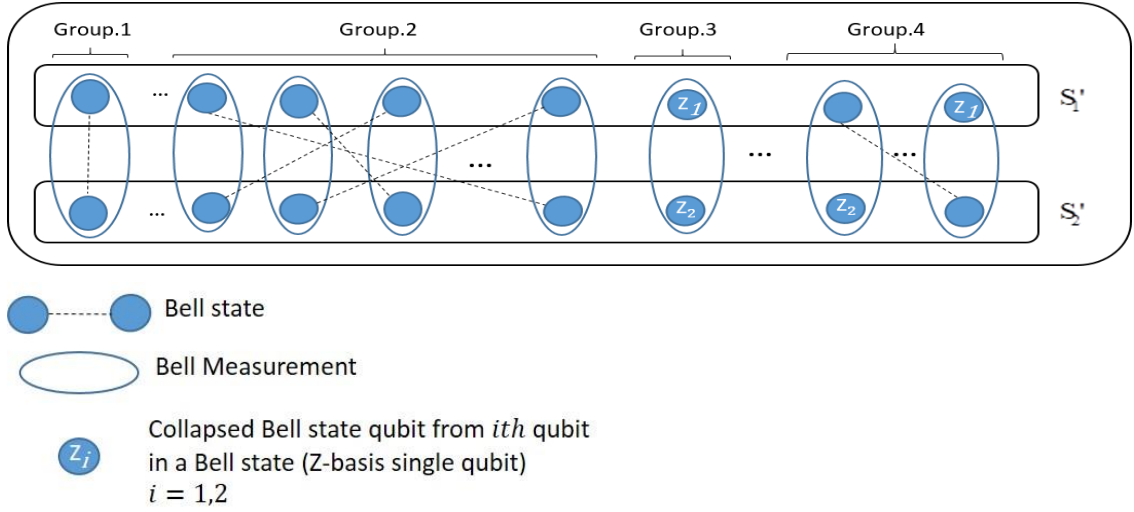

Figure.1 illustration of the Entanglement Correlations among collapsed Bell state qubits and reordered Bell states

In Group 1 and Group 2, if the Bell states in $\{|\phi^+\rangle,|\phi^-\rangle,|\psi^+\rangle,|\psi^-\rangle\}$ are encoded respectively into two-bit classical information "00", "01", "10", "11", then the following properties, which will be useful in the design of the proposed protocol, can be derived:

**Group 1. an original Bell state with a Bell Measurement**

Consider a Bell state $|IS\rangle$ in $\{|\phi^+\rangle,|\phi^-\rangle,|\psi^+\rangle,|\psi^-\rangle\}$. While a Bell Measurement is performed on $|IS\rangle$, one can get a Bell Measurement result $MR$ in $\{|\phi^+\rangle,|\phi^-\rangle,|\psi^+\rangle,|\psi^-\rangle\}$, where the $MR$ should obey the following equation.

$$MR=IS \qquad (1)$$

**Group 2. Entanglement Swapping among reordered Bell states [31,32]**

Consider $k$ Bell states $|IS_i\rangle$ in $\{|\phi^+\rangle,|\phi^-\rangle,|\psi^+\rangle,|\psi^-\rangle\}$, $1\leq i \leq k$. While the multiple Bell Measurements are performed on pairs of qubits from different Bell states,

one can get $k$ Bell Measurement results $|MR_i\rangle$ in $\{|\phi^+\rangle,|\phi^-\rangle,|\psi^+\rangle,|\psi^-\rangle\}$, $1\leq i \leq k$, where the $|MR_i\rangle$ should obey the following equation.

$$MR_1 \oplus .. \oplus MR_k = IS_1 \oplus .. \oplus IS_k \qquad (2),$$

where $\oplus$ is the exclusive-or operation; $k$ is a positive integer indicating the number of Bell states, and $k \in Z^+$.

Next, in Group 3 and Group 4, for deriving the Entanglement correlations in the collapsed Bell state qubits and original Bell states, which will be useful in the proposed protocol, the Bell states $\{|\psi^\pm\rangle, |\phi^\pm\rangle\}$ will be encoded respectively into one bit classical information "1" and "0" (as shown in Table.1).

The Bell states can be denoted as $|\phi^\pm\rangle = \frac{1}{\sqrt{2}}(|00\rangle \pm |11\rangle)$, $|\psi^\pm\rangle = \frac{1}{\sqrt{2}}(|01\rangle \pm |10\rangle)$, and the two single photons in Z-basis can be presented by two Bell states described below.

$$\begin{aligned}
|00\rangle &= \frac{1}{\sqrt{2}}(|\phi^+\rangle + |\phi^-\rangle) \\
|01\rangle &= \frac{1}{\sqrt{2}}(|\psi^+\rangle + |\psi^-\rangle) \\
|10\rangle &= \frac{1}{\sqrt{2}}(|\psi^+\rangle - |\psi^-\rangle) \\
|11\rangle &= \frac{1}{\sqrt{2}}(|\phi^+\rangle - |\phi^-\rangle)
\end{aligned} \qquad (3)$$

Table.1 the encoding for Group 3 and Group 4

| Bell state | Classical bit |
|---|---|
| $|\psi^{\pm}\rangle$ | 1 |
| $|\phi^{\pm}\rangle$ | 0 |

**Group 3. Entanglement Correlation among two collapsed Bell state qubits**

Consider two original Bell states $|IS_1\rangle, |IS_2\rangle$ in $\{|\psi^{\pm}\rangle, |\phi^{\pm}\rangle\}$, based on the Table 1, the $|IS_1\rangle, |IS_2\rangle$ will be encoded respectively into two one-bit classical information $is_1$, $is_2$ in $\{0,1\}$. If one qubit in each Bell state is measured in Z-basis with the Z-basis measurement results, $z\_mr_1$ and $z\_mr_2$, respectively, then the remaining qubit in each Bell state will become a collapsed Bell state qubit (i.e. Z-basis single photon) $|z_1\rangle, |z_2\rangle$ in $\{|0\rangle, |1\rangle\}$. When a Bell Measurement is performed on the two collapsed Bell state qubits $|z_1\rangle$ and $|z_2\rangle$, based on Eq.(3), one can obtain the Bell Measurement result $MR$ in $\{|\psi^{\pm}\rangle, |\phi^{\pm}\rangle\}$ which can also be encoded into one bit classical information $mr$ in $\{0,1\}$, based on Table 1. In this case, we can derive the following equation.

$$z\_mr_{2(1)} = \sigma_{is_{2(1)}} \sigma_{mr} \left( z\_mr_{1(2)} \right), \begin{cases} if\ \sigma_0,\ do\ nothing \\ if\ \sigma_1,\ perform\ \neg\ operation \end{cases} \quad (4).$$

For example, the two Bell states $|IS_1\rangle = |IS_2\rangle = |\phi^+\rangle$ can be encoded as $is_1 = is_2 = 0$ based on Table 1. If $z\_mr_1 = 0$, $z\_mr_2 = 1$ are obtained respectively by measuring one qubit in $|IS_1\rangle$, $|IS_2\rangle$, then the collapsed qubits in both Bell states will be $|0\rangle$ and $|1\rangle$, respectively. According to Eq.(3), when a Bell Measurement is performed on ($|0\rangle, |1\rangle$), the Bell Measurement result $MR$ could be either $|\psi^+\rangle$ or $|\psi^-\rangle$, which will be encoded as $mr = 1$ by Table 1. Consequently, we have

$0 = z\_mr_1 = \neg(z\_mr_2)$ or $1 = z\_mr_2 = \neg(z\_mr_1)$.

## Group 4. Entanglement Correlations among collapsed Bell state qubits and reordered Bell states

To better understand this group, let us consider the following example. Assume after Bell Measurements are performed on qubits pairs from $S_1'$ and $S_2'$ as described earlier in this section, we have some entanglement correlations like the one described in Figure 2. There are four collapsed Bell state qubits inserted in reordered Bell states (see also Figure 2). After the Bell Measurements are performed on the qubit pairs in order, based on the entanglement correlation, one can obtain two entanglement groups Group 4-1 and Group 4-2. Obviously, Group 4-1 is belonging to Group 4 as shown in Figure 1 with two collapsed Bell state qubits and one original Bell state. On the other hand, Group 4-2 is not in the same form as Group 4 in Figure 1. However, after the certain arrangement, where the collapsed Bell state qubits can be moved to both end of Group 4-2 and the rest of the Bell state qubits can be permuted as shown in Figure 2. The equation for Group 4 can be derived in the following.

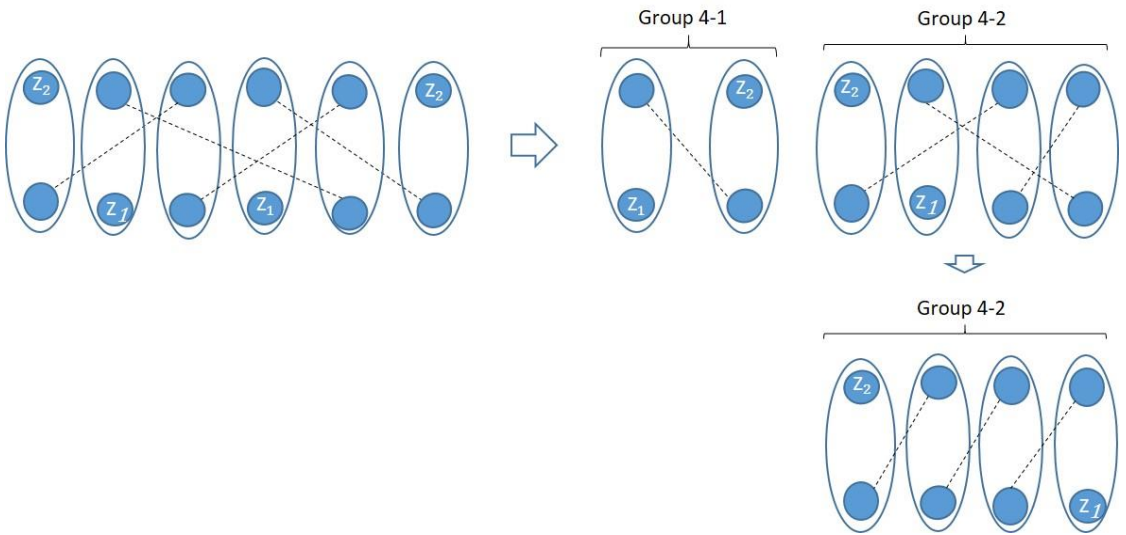

Figure 2. The possible permutation between collapsed Bell state qubits and Bell states

Consider two collapsed Bell state qubits $|z_i\rangle$ in $\{|0\rangle, |1\rangle\}$, $i = 1, 2$ are obtained from the two corresponding Z-basis measurement results $z\_mr_i$, on two Bell states $|IS_i\rangle$ in $\{|\psi^\pm\rangle, |\phi^\pm\rangle\}$, $i=1,2$, and another $j$ reordered Bell states $|IS_m\rangle$ in $\{|\psi^\pm\rangle, |\phi^\pm\rangle\}$, $1 \leq m \leq j$. If the $j$ reordered Bell states $|IS_m\rangle$ are in-between ($|z_1\rangle, |z_2\rangle$), then the Entanglement Correlations will occur while performing the Bell Measurements, and one will obtain $j+1$ Bell Measurement results $MR_y$ in $\{|\psi^\pm\rangle, |\phi^\pm\rangle\}$ which will be encoded as $mr_y$ based on Table 1, $1 \leq y \leq j+1$. Then the following equation can be satisfied.

$$z\_mr_{2(1)} = \sigma_{is_{2(1)}} \prod_{y=1}^{j+1} \sigma_{mr_y} \left(z\_mr_{1(2)}\right), \begin{cases} if\ \sigma_0,\ do\ nothing \\ if\ \sigma_1,\ perform\ \neg\ operation \end{cases} \quad (5),$$

For example, if the two Z-basis measurement results, $z\_mr_1=0$ and $z\_mr_2=1$ obtained respectively from $|IS_i\rangle = |\phi^+\rangle$ encoded as $is_i=0$, $i=1,2$, then the two collapsed Bell state qubits $|z_i\rangle$, $i=1,2$, are $|0\rangle$ and $|1\rangle$, respectively. For $j=1$, and $|IS_m\rangle = |\phi^+\rangle$. When the Entanglement Correlation occurs between ($|z_1\rangle, |z_2\rangle$) and $|IS_m\rangle$ by performing the Bell Measurements, the Bell Measurement results $MR_1, MR_2$ will be $|\phi^+\rangle|\psi^+\rangle$, $|\phi^+\rangle|\psi^-\rangle$, $|\phi^-\rangle|\psi^+\rangle$, $|\phi^-\rangle|\psi^-\rangle$, $|\psi^+\rangle|\phi^+\rangle$, $|\psi^+\rangle|\phi^-\rangle$, $|\psi^-\rangle|\phi^+\rangle$ or $|\psi^-\rangle|\phi^-\rangle$, where the $MR_1, MR_2$ would be encoded as $mr_1$, $mr_2$, then we have

$1 = z\_mr_2 = \neg(z\_mr_2)$ or $0 = z\_mr_1 = \neg(z\_mr_2)$

### 3. Proposed scheme

A new Mediated RSQKD protocol is proposed here, in which two classical participants, Alice and Bob, want to share a secret key with the assistance of an untrusted TP. The assumptions of the proposed scheme are shown as follows:

- The quantum channels share between TP and the two classical participants are

- ideal (i.e. non-lossy and noiseless).
- There are public classical channels shared between TP and both classical participants.
- Alice and Bob have an authenticated classical channel between each other. Anyone can obtain information from the channel, but he/she cannot modify the information.
- TP is a quantum user who have complete quantum capabilities. In this proposed protocol, TP can perform any attack.
- Alice and Bob are classical users who only have limited quantum capabilities including Z-basis measurement, reflecting the qubits, and reordering the qubits.

The steps of the proposed Mediated RSQKD protocol are illustrated as follows (see Figure.2):

**Step 1.** TP generates $n$ Bell states $\left|\phi^+\right\rangle_i, 1 \leq i \leq n$, and takes all the 1st particles (2nd particles) of the Bell state to form ordered qubit sequence $S_1 = \{q_1^1, q_1^2, ..., q_1^n\}$ ($S_2 = \{q_2^1, q_2^2, ..., q_2^n\}$). Then $S_1$ and $S_2$ are sent to Alice and Bob, respectively. Here, the number of the Bell-states $n$, $n \in Z^+$, is a variable which can be selected by the participants according to the demanded key length and detection rate.

**Step2.** When Alice (Bob) receives $S_1$ ($S_2$), she (he) randomly chooses $\frac{n}{2}$ qubits in $S_1$ ($S_2$) and performs Z-basis measurement on them with the Z-basis measurement results denoted respectively as $z\_mr_1^{a_i}$ ($z\_mr_2^{b_j}$), where $a_i$ ($b_j$) indicates the position of the measured qubit in $S_1$ ($S_2$), $1 \leq a_i, b_j \leq n$, $1 \leq i, j \leq \frac{n}{2}$, and the corresponding collapsed Bell state qubits are denoted as $\left|z_2^{a_i}\right\rangle$ ($\left|z_1^{b_j}\right\rangle$). Afterward, the remaining qubits are reordered to form the qubit sequence $Q_1$ ($Q_2$). After that, Alice (Bob) sends $Q_1$ ($Q_2$) to TP.

**Step3.** When TP receives $Q_1$ and $Q_2$ from Alice and Bob, TP performs Bell Measurement on the *kth* particle pair (i.e., the *kth* particle from $Q_1$ and the *kth* particle from $Q_2$ are performed Bell Measurement, $1 \leq k \leq \frac{n}{2}$). Afterward, TP obtains the measurement results **MR**={ $MR_1$,..., $MR_{\frac{n}{2}}$ }, and sends **MR** to Alice and Bob via public classical channels.

**Step4.** After Alice (Bob) receives the **MR** from TP, Alice (Bob) publishes the order of $Q_1$ ($Q_2$) to each other via an authenticated classical channel. Then, Alice (Bob) can identify the entanglement correlations of the particles. The possible correlations are illustrated in Figure.1 and the possible relationship in measurement results is described as follows:

**Case1.** If Alice (Bob) performs Z-basis measurement in Step 2 with the Z-basis measurement result $z\_mr_1^{a_i}$ ($z\_mr_2^{b_j}$) where $a_i = b_j$, for some $i, j$, then $z\_mr_1^{a_i} = z\_mr_2^{b_j}$ and the measurement result will be treated as a raw key bit.

**Case2.** TP performs Bell Measurements on the Bell state qubits $(q_1^i, q_2^j)$. If $i = j$, it is a pair of original Bell state and the Bell Measurement result should follow Eq.(1). If $i \neq j$, it is a pair of qubits from different Bell states and the Bell Measurement result should follow Eq.(2); otherwise Alice and Bob will abort the protocol and restart from the beginning (see also Group.1 and Group.2 in Figure.1).

**Case3.** If TP performs Bell Measurement on a pair of particles formed by two collapsed Bell state qubits ($z_1^{b_j}, z_2^{a_i}$), the Bell Measurement result should follow Eq.(4), and the corresponding $z\_mr_1^{a_i}$ will be treated as a raw key bit (see also Group.3).

**Case4.** If Entanglement Correlation occurs among reordered Bell states and ($z_1^{b_j}, z_2^{a_i}$), then Alice (Bob) discloses the corresponding $z\_mr_2^{b_j}$ ($z\_mr_1^{a_i}$), and the Bell Measurement results should follow Eq.(5); otherwise Alice and Bob will abort the protocol and restart from the beginning. (see also Group 4 in Figure.1)

**Step5.** Alice and Bob will perform the privacy amplification [45,46] on the raw key bits to form the session key.

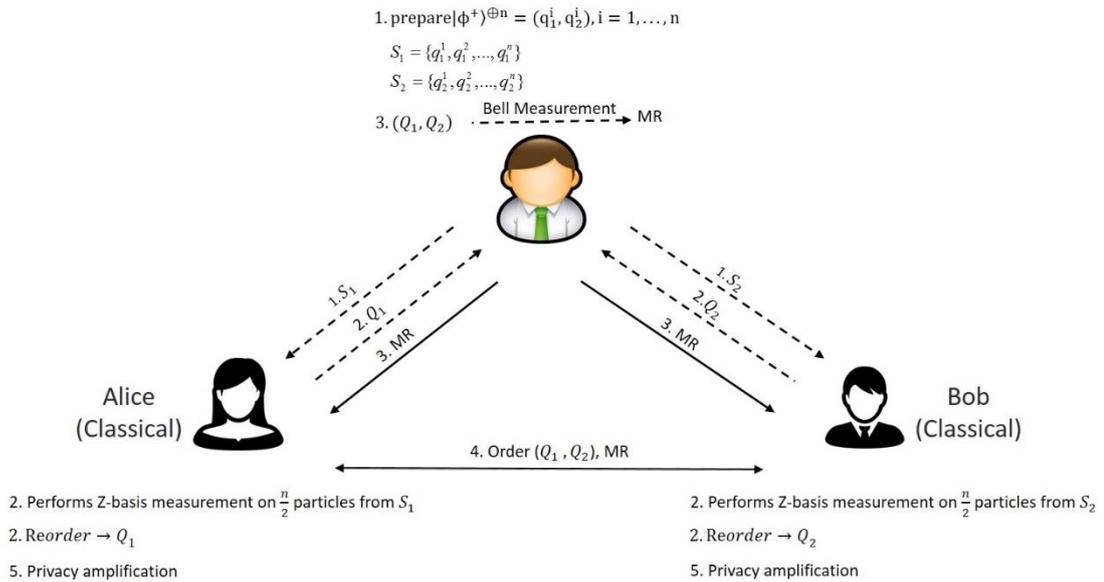

Figure.2 The proposed scheme

## 4. Security analyses and comparisons

The security analysis for the general attacks and the comparison with the existing SQKD protocols are described in Section 4.1 and Section 4.2, respectively.

**4.1 Security analyses**

Because it is much easier for TP to attack the SQKD protocol than outsiders (e.g., TP can send fake initial states, disclose the wrong measurement result…or so on), the security analyses of this study focus on TP's attacks. In a word, if the proposed protocol can resist the TP's attacks, then the proposed protocol can resist the same attacks from outsiders. Moreover, because TP is assumed to be an untrusted one in the proposed SQKD protocol, TP can perform any attacks and try to obtain the session key shared with Alice and Bob.

*Measurement attack*

For obtaining the information of the session key in Step 3, while TP receives $Q_1$ and $Q_2$, he/she may perform Z-basis measurements instead of Bell Measurements, and publish fake Bell Measurement results. In this way, if TP passes the detection, he can obtain the $z\_mr_2^{b_j}$ and $z\_mr_1^{a_i}$ in Case 3, which are raw key bits, by performing the Z-basis measurement on the corresponding collapsed Bell state qubit $z_1^{b_j}, z_2^{a_i}$ in $Q_1$ and $Q_2$. <span style="color:red">However, the Case 2 in Step 4 and the Case 4 in Step 4 can detect this attack</span>. The probability of the detecting the Case 2, is $\frac{1}{4}$, and the probability for the fake Bell measurement results to follow the Eq.(1) and the Eq.(2) is $\frac{1}{2}$. Therefore, in this case, the detection rate of this attack is $\frac{1}{4} * \frac{1}{2} = \frac{1}{8}$.

In the Case 4, Alice (Bob) discloses the <span style="color:red">corresponding</span> $z\_mr_2^{b_j}$ ($z\_mr_1^{a_i}$), and the fake Bell Measurement results which TP forges should follow Eq.(5). The probability of Case 4 is $\frac{1}{2}$, and the probability for the fake Bell measurement result to be detected on each bit of session key is $\frac{1}{2}$. The detection rate for this attack in this case is $\frac{1}{2} * \frac{1}{2} = \frac{1}{4}$. The probability of Eve to simultaneously pass the detection of the Case 2 and Case 4 is $\frac{7}{8} * \frac{3}{4} = \frac{21}{32}$. Hence, the total detecting probability of the measurement attack of TP is $1-(\frac{21}{32})^t$, $t$ is the number of the shared secret key bits, $t \in Z^+$. When $t$ is large enough, the detection rate is approximately 100%.

*Modification attack*

In the modification attack, TP intends to let Alice and Bob share an inconsistent session key by modifying their Z-basis measurement result $z\_mr_1^{a_i}$ and $z\_mr_2^{b_j}$ obtained in Step 2. For example, in Step 1, while TP transmits the ordered qubit sequence $S_1$ and $S_2$ respectively to Alice and Bob, he/she performs arbitrary unitary operation in $\{\sigma_z, \sigma_x, i\sigma_y, H\}$ on all of the particles in $S_1$ to obtain $S_1^E$. Then $S_1^E$ and $S_2$ are sent respectively to Alice and Bob. After Alice (Bob) executes the legal operations in Step 2, and sends the $Q_1^E$ ($Q_2$) back to TP. TP performs the same kind of unitary operations chosen in Step 1 on all of the particles in $Q_1^E$ to obtain $Q_1$. Then he/she performs the Bell Measurements as usual. Although executing the same unitary operation twice on an Bell state does not change the initial state and destroy the entangled relation, the $z\_mr_1^{a_i}$ and $z\_mr_2^{b_j}$ will result in inconsistent values because the measured qubits in $S_1$ and $S_2$ were performed the unitary operation once.

However, this attack can be avoided by the Case 4 in Step 4 (see also Group.4 in Figure.1). The corresponding $z\_mr_1^{a_i}$ and $z\_mr_2^{b_j}$ in Case 4 disclosed by Alice and Bob should follow Eq.(6), and the probability of the Case 4 is $\frac{1}{2}$. If TP modifies m bits, $1 \leq m \leq N$, where $N$ is the length of the session key, then the detection probability for this attack is $1-\left(\frac{1}{2}\right)^m$. On the other hand, if TP modified only a particular qubit, it has the probability of $\frac{1}{2}$ to be treated as a raw key bit. That is, if TP wants to modify a certain raw key bit successfully, he/she needs to modify the qubits as many as possible. However, if the modified qubits number $m$ is large enough, the detection rate will be close to 1.

*Trojan-horse attack [33-36]*

The attacker could also try to generate the Trojan-horse photons where the wavelength is not sensitive to users' devices, and then insert these photons into the $S_1$ and $S_2$ in Step 1. After the operations in Step 2, TP can obtain the reorder of $Q_1$ ($Q_2$) by recognizing the wavelength of each Trojan-horse photons before Step 4. Fortunately, this attack can be avoided by utilizing the photon number splitter and the optical wavelength filter devices [37-39] to filter out the Trojan-horse photons.

*Collective attack*

Collective attack is a general attack where the attacker entangles the ancillary particles with the original Bell states for acquiring useful information. We give an analysis to show that our protocol can resist this attack [42-44].

In the proposed protocol, TP may perform the unitary operation $U$ where $U*U=I$, on $q_1^i$ in $S_1$, $1 \leq i \leq n$, to entangle the particles with the ancillary particles $E = \{|E_1\rangle, |E_2\rangle, ..., |E_n\rangle\}$, then the state of $q_1^i$ will be mapped into the ancillary qubits. After the operations in Step 2 and Step 3, Alice (Bob) will announce the reorder of $Q_1$ ($Q_2$) in Step 4, then TP can know the positions which Alice (Bob) performs Z-basis measurement, and further obtain the raw key bits by measuring the ancillary particles in the corresponding positions. The analysis and the operation $U$ are described as follows:

$$
\begin{aligned}
U|00\rangle \otimes |E_i\rangle &= a_0|00\rangle|e_0\rangle + a_1|01\rangle|e_1\rangle + a_2|10\rangle|e_2\rangle + a_3|11\rangle|e_3\rangle \\
U|01\rangle \otimes |E_i\rangle &= b_0|00\rangle|f_0\rangle + b_1|01\rangle|f_1\rangle + b_2|10\rangle|f_2\rangle + b_3|11\rangle|f_3\rangle \\
U|10\rangle \otimes |E_i\rangle &= c_0|00\rangle|g_0\rangle + c_1|01\rangle|g_1\rangle + c_2|10\rangle|g_2\rangle + c_3|11\rangle|g_3\rangle \\
U|11\rangle \otimes |E_i\rangle &= d_0|00\rangle|h_0\rangle + d_1|01\rangle|h_1\rangle + d_2|10\rangle|h_2\rangle + d_3|11\rangle|h_3\rangle
\end{aligned} \quad (6),
$$

where $|E_i\rangle$ denotes the states which the TP prepares, and $|e_0\rangle, |e_1\rangle, |e_2\rangle, |e_3\rangle, |f_0\rangle, \ldots, |h_3\rangle$ are the states changed from the states $|E_i\rangle$ after performing the operation $U$, in which $|a_0|^2 + |a_1|^2 + |a_2|^2 + |a_3|^2 = 1$, $|b_0|^2 + |b_1|^2 + |b_2|^2 + |b_3|^2 = 1$, $|c_0|^2 + |c_1|^2 + |c_2|^2 + |c_3|^2 = 1$, $|d_0|^2 + |d_1|^2 + |d_2|^2 + |d_3|^2 = 1$.

Give an example which is shown as follows:

$$
\begin{aligned}
U|\phi^+\rangle \otimes |E_i\rangle &= \frac{1}{\sqrt{2}}(U|00\rangle \otimes |E_i\rangle + U|11\rangle \otimes |E_i\rangle) \\
&= \frac{1}{2}\begin{bmatrix}
|\phi^+\rangle(b_0|f_0\rangle + b_3|f_3\rangle + c_0|g_0\rangle + c_3|g_3\rangle) \\
+ |\phi^-\rangle(b_0|f_0\rangle - b_3|f_3\rangle + c_0|g_0\rangle - c_3|g_3\rangle) \\
+ |\psi^+\rangle(b_1|f_1\rangle + b_2|f_2\rangle + c_1|g_1\rangle + c_2|g_2\rangle) \\
+ |\psi^-\rangle(b_1|f_1\rangle - b_2|f_2\rangle + c_1|g_1\rangle - c_2|g_2\rangle)
\end{bmatrix}
\end{aligned} \quad (7)
$$

Assuming there is a Bell state $|\phi^+\rangle = \frac{1}{\sqrt{2}}(|00\rangle + |11\rangle)$, TP performs the operation $U$ to entangle the ancillary particle with the original Bell state. For passing the detection, TP will eliminate the following incorrect items: $|\phi^-\rangle(b_0|f_0\rangle - b_3|f_3\rangle + c_0|g_0\rangle - c_3|g_3\rangle)$, $|\psi^+\rangle(b_1|f_1\rangle + b_2|f_2\rangle + c_1|g_1\rangle + c_2|g_2\rangle)$, $|\psi^-\rangle(b_1|f_1\rangle - b_2|f_2\rangle + c_1|g_1\rangle - c_2|g_2\rangle)$. Consequently, TP needs to set $b_0|f_0\rangle - b_3|f_3\rangle + c_0|g_0\rangle - c_3|g_3\rangle = b_1|f_1\rangle + b_2|f_2\rangle + c_1|g_1\rangle + c_2|g_2\rangle = b_1|f_1\rangle - b_2|f_2\rangle + c_1|g_1\rangle - c_2|g_2\rangle = \bar{0}$, where $\bar{0}$ denotes zero vector. If TP accomplishes the above-mentioned operations, then TP will

not be detected by classical participants. Nevertheless, because $b_3|f_3\rangle = c_3|g_3\rangle$ and $b_0|f_0\rangle = c_0|g_0\rangle$, TP cannot distinguish between $b_3|f_3\rangle$ and $c_3|g_3\rangle$, $b_0|f_0\rangle$ and $c_0|g_0\rangle$. In other words, TP cannot obtain any information without being detected.

**4.2 Comparisons**

This subsection compares the proposed SQKD protocol to the other existing SQKD protocols which are also three-party SQKD with untrusted TP (see also Table 2).

The existing SQKD protocols, Krawec's scheme [29] and Liu et al.'s scheme [30], are designed in the Measure-Resend environment and the Measurement-Free environment, respectively. However, our proposed scheme is designed in the Randomization-Based environment. In terms of the quantum resource, in Krawec's scheme [29] and Liu et al.'s scheme [30], TP needs to generate Bell states and the classical participants have to generate single photons, on the other hand, in our proposed scheme, TP has to prepare Bell states, but the classical participants don't need to prepare any photons.

Next, the qubit efficiency is defined by the following formula [40,41]:

$$QE = \frac{b_s}{q_t} \quad (8)$$

, where The paraeter $b_s$ denotes the number of bits of the shared session key, and the parameter $q_t$ denotes the number of the total particles used in the protocol. Based on the Eq.(6), for $3m$ shared session key bits, the proposed protocol need to generate $8m$ pairs of Bell States, $m \in Z^+$, the qubit efficiency of the proposed MRSQKD scheme is $\frac{3}{16}$. On the other hand, and the qubit efficiencies of Krawec's scheme [29] and Liu's scheme [30] are $\frac{1}{32}$ and $\frac{1}{8}$, respectively.

Table 2. Comparison with other existing mediated SQKD protocols

|  | **Krawec's SQKD [29]** | **Liu et al.'s SQKD [30]** | **Proposed SQKD** |
|---|---|---|---|
| Semi-quantum environment | Measure-resend | Measurement-free | Randomization-Based |
| Quantum resources | (1) Single photon (2) Bell states | (1) Single photon (2) Bell states | Bell state |
| Qubit efficiency | 1/32 | 1/8 | 3/16 |

## 5. Conclusions

This study uses the entanglement Swapping between reordered Bell states and the entanglement correlations among collapsed Bell state qubits and reordered Bell states to propose a semi-quantum key distribution protocol allowing two classical participants to share a session key with the assistance of an untrusted TP (quantum server). The study also shows that the proposed SQKD protocol can resist the malicious behavior of the untrusted TP or other attackers.

The performance of the proposed protocol is better than Krawec's SQKD protocol and Liu's SQKD protocol designed in Measure-Resend environment and Measurement-Free environment respectively. The security analyses described in Section 4 shows that our proposed RSQKD protocol is free from various well-known attacks.

The purpose of "Semi-Quantum" is to reduce the quantum burden of the classical participants. Hence, how to further decrease the quantum burden of the classical participant will be an interesting and practical future research. We design a semi-quantum cryptographic protocols with more fewer quantum capabilities some day in the future.

**Acknowledgment**

This research is partially supported by the Ministry of Science and Technology, Taiwan, R.O.C, under the Contract No. MOST 107-2221-E-006 -077 -

This research is partially supported by the Ministry of Science and Technology, Taiwan, R.O.C, under the Contract No. MOST 107-2221-E-006 -077 -